\documentclass[final,3p,11pt]{elsarticle}

\makeatletter
\def\ps@pprintTitle{%
 \let\@oddhead\@empty
 \let\@evenhead\@empty
 \def\@oddfoot{%
 \footnotesize\itshape
       Preprint accepted by \ifx\@journal\@empty Elsevier
       \else\@journal\fi\hfill%
       August 24, 2020
 }
 \let\@evenfoot\@oddfoot}
\makeatother

\usepackage{amssymb}

\usepackage[htt]{hyphenat}

\usepackage{microtype}
\usepackage[T1]{fontenc}
\usepackage[lcgreekalpha]{stix}

\usepackage{color}

\usepackage{listings}
\lstdefinestyle{mystyle}{
	basicstyle=\ttfamily\small,
	breakatwhitespace=true,
	breaklines=true,
}
\usepackage{hyperref}
\hypersetup{
  colorlinks
}

\newcounter{bla}

\journal{Computer Physics Communications}

\begin{document}

\begin{frontmatter}



\title{TurboPy: A Lightweight Python Framework for Computational Physics}


\author[NRL]{A.~S.~Richardson\corref{author}}
\author[NRL]{D.~F.~Gordon}
\author[NRL]{S.~B.~Swanekamp}
\author[NRL]{I.~M.~Rittersdorf}
\author[NRL]{P.~E.~Adamson}

\author[syntek]{{O.~S.~Grannis}}
\author[syntek]{{G.~T.~Morgan}}
\author[syntek]{{A.~Ostenfeld}}
\author[syntek]{{K.~L.~Phlips}}
\author[syntek]{{C.~G.~Sun}}
\author[syntek]{{G.~Tang}}
\author[syntek]{{D.~J.~Watkins}}

\cortext[author] {Corresponding author.\\\textit{E-mail address:} \href{mailto:steve.richardson@nrl.navy.mil}{steve.richardson@nrl.navy.mil}}
\address[NRL]{U.S.~Naval Research Laboratory, Plasma Physics Division, Washington, DC}
\address[syntek]{{Syntek Technologies, Fairfax, VA}}

\tnotetext[dist]{DISTRIBUTION A. Approved for public release: distribution unlimited. \\
\copyright~2020. This manuscript version is made available under the CC-BY-NC-ND 4.0 license 
\url{http://creativecommons.org/licenses/by-nc-nd/4.0/}
}

\begin{abstract}
Computational physics problems often have a common set of aspects to them that any particular numerical code will have to address. Because these aspects are common to many problems, having a framework already designed and ready to use will not only speed the development of new codes, but also enhance compatibility between codes. 

Some of the most common aspects of computational physics problems are: a grid, a clock which tracks the flow of the simulation, and a set of models describing the dynamics of various quantities on the grid. Having a framework that could deal with these basic aspects of the simulation in a common way could provide great value to computational scientists by solving various numerical and class design issues that routinely arise.

This paper describes the newly developed computational framework that we have built for rapidly prototyping new physics codes. This framework, called turboPy, is a lightweight physics modeling framework based on the design of the particle-in-cell code turboWAVE. It implements a class (called {\tt Simulation}) which drives the simulation and manages communication between physics modules, a class (called {\tt PhysicsModule}) which handles the details of the dynamics of the various parts of the problem, and some additional classes such as a {\tt Grid} class and a {\tt Diagnostic} class to handle various ancillary issues that commonly arise.
\end{abstract}


\end{frontmatter}



{\bf PROGRAM SUMMARY}

\begin{small}
\noindent
{\em Program Title:}  TurboPy \\
{{\em Developer's repository link:} \url{https://github.com/NRL-Plasma-Physics-Division/turbopy}}\\
{\em Licensing provisions: CC0 1.0 }                               \\
{\em Programming language:}         Python           \\
{\em Nature of problem:}\\
 Many computation physics problems have a common set of aspects to them that are often addressed in a custom way in every different code, which leads to lengthy and redundant development and testing, as well as introducing roadblocks to interoperability. \\
{\em Solution method:}\\
Implement a set of python classes as a lightweight framework that deals with these common problems, so that development time on new computational physics codes is reduced, and interoperability and reusability are increased. \\
{{\em References:} A.~S.~Richardson et al., TurboPy: A lightweight computational physics framework. NRL-Plasma-Physics-Division/turbopy (v2020.08.05). doi:10.5281/zenodo.3973693}
\end{small}

\section{Introduction}
\label{intro}

The development of a new computational framework for physics simulations is motivated by two main factors. The first motivation is to increase the productivity of researchers in scientific computing by providing a pre-built framework for dealing with numerical and code design issues that are common to many types of computational physics workflows. In addition to saving on development time, this also provides a starting point for new codes that encourages the developers to adhere more closely to the best practices for scientific computing\cite{10.1371/journal.pbio.1001745}. An example of a similarly motivated framework is NumPy, which provides common tools to address problems involving vector and matrix math\cite{numpy,5725236}. 

The second motivation for developing a new framework is to provide a streamlined path from implementing and testing new algorithms (for which a high-level language like python has many advantages) to the implementation of the new algorithms as C++ modules for use with turboWAVE\cite{turboWAVE,893300}. This is achieved in turboPy by mirroring the class design used in turboWAVE. This helps ensure that new turboPy modules will translate easily into modules for turboWAVE. 

In order to maximize the flexibility of turboPy{\cite{turbopy}}, it was designed to be a lightweight framework which provides valuable tools for any computational workflow that requires a grid and clock. There are a wide range of computational physics problems that exhibit these features. These problems range from full plasma simulation codes to simple scripts for post processing the outputs of other simulations. 

In Sec.~\ref{design} we describe the class design used in turboPy. 
Then in Sec.~\ref{turbopy} we describe some details of the implementation of this structure in python, and describe how to use turboPy by subclassing the appropriate abstract base classes. 
A few concepts for potential use cases for the turboPy framework are then described in Sec.~\ref{example}. Conclusions and thoughts on future directions this framework could take are in Sec.~\ref{con}.

\section{TurboPy class design}
\label{design}
The main set of classes that form the core architecture of turboPy are: \texttt{Simulation}, \texttt{PhysicsModule}, \texttt{ComputeTool}, and \texttt{Diagnostic}. An instance of the \texttt{Simulation} class owns the various objects that define the simulation, and drives the simulation by coordinating the objects. This class is not meant to be subclassed. The \texttt{PhysicsModule}, \texttt{ComputeTool}, and \texttt{Diagnostic} classes are subclassed, however, in order to define the specific behavior that is needed in the simulation. Instances of {\tt PhysicsModule} subclasses define the physics that occurs on each time step. Instances of {\tt ComputeTool} subclasses define numerical methods that physics modules may need to do their work. Instances of {\tt Diagnostic} subclasses define the desired output quantities.

\begin{figure}[htbp]
\begin{center}
\includegraphics[width=12cm]{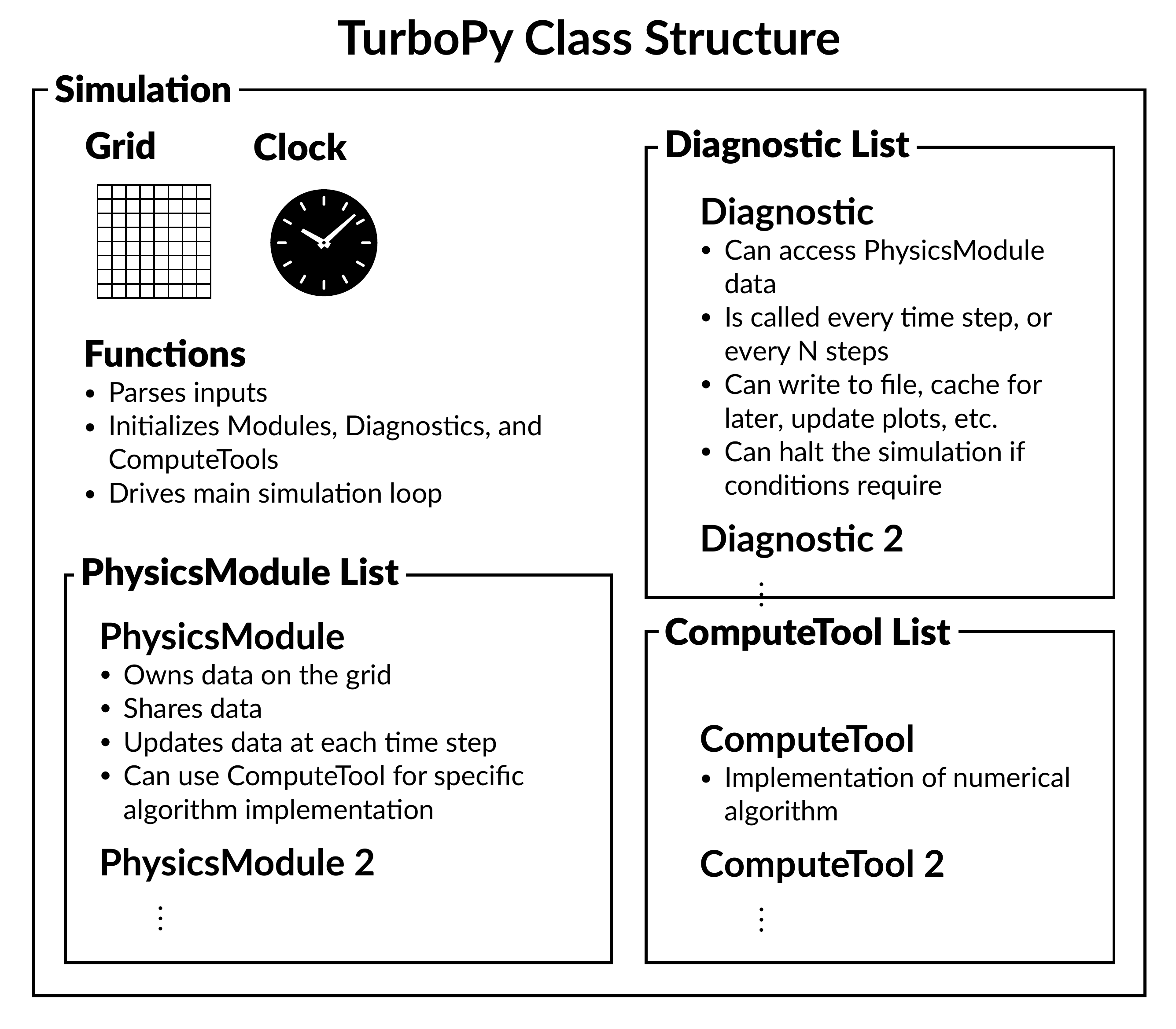}
\caption{Structure of a turboPy simulation, showing the hierarchy of objects that make up the simulation.}
\label{structure}
\end{center}
\end{figure}

\subsection{Brief description of the core turboPy classes}
\subsubsection{The Simulation class}
The \texttt{Simulation} class is the main class. Every time turboPy runs, an instance of this class is created, which handles the setup of the problem, running the main simulation loop, and coordinating the physics modules and diagnostics. 

Member Variables: The simulation class has several python lists, which hold the physics modules (\texttt{PhysicsModule} objects), the compute tools (\texttt{ComputeTool} objects), and the diagnostics (\texttt{Diagnostic} objects). When a user sets up a turboPy simulation, they must describe the various physics modules, compute tools, and diagnostics they need in their simulation. The appropriate objects are created at runtime, and added to the appropriate lists in the \texttt{Simulation} instance.

Member Functions: The main subroutine within the \texttt{Simulation} class is \texttt{run}. Calling \texttt{run} performs two main functions. First, it calls \texttt{prepare\_simulation}, which creates and initializes the objects (including the grid, clock, physics modules, and diagnostics) needed for the simulation. These objects are specified in the python dictionary \texttt{input\_data} that is passed to the constructor for the \texttt{Simulation} object. {Alternatively, a helper function \texttt{construct\_simulation\_from\_toml} was written which loads a specified \texttt{TOML} formatted file\cite{toml}, converts it to a python dictionary, and returns a \texttt{Simulation} object created from that dictionary.}
The function \texttt{run} then drives the main simulation loop, calling \texttt{fundamental\_cycle} at each time step. Each call to \texttt{fundamental\_cycle} performs the sequence of steps that need to happen at every time step. The first step is looping over all the diagnostics, allowing them to perform their work as appropriate. Then the list of physics modules is looped over once, calling \texttt{reset} on every physics module. The list is then looped over a second time, calling \texttt{update} on each physics module. The \texttt{update} call is where the numerical work needed by each physics model happens. Finally, the clock is advanced.

\subsubsection{The PhysicsModule base class}
TurboPy uses the concept of physics modules to enable extendability of computational physics codes written within the framework. The abstract class \texttt{PhysicsModule} is used as a base class for all physics modules, which allows them to interact with each other and the core turboPy code without needing to know each other's internal details. This modularity provides a loose coupling between different sections of code, which greatly enhances code readability, maintainability, and extendability. This design is based on modern object-oriented programming best practices, and makes applications written within the turboPy framework very flexible.

Each subclass of the \texttt{PhysicsModule} class must implement its own custom \texttt{reset} and \texttt{update} functions. This is where the numerical work that needs to happen at each timestep is defined. Additionally, it can optionally implement custom \texttt{inspect\_resource} and \texttt{exchange\_resources} functions. The \texttt{exchange\_resources} function is the one that tells other physics modules about data provided by this particular module, while the \texttt{inspect\_resource} function is the one where this physics module gets to ``inspect'' the resources provided by other modules to see if they provide needed data. Additional details about how these functions work is provided in Sec.~\ref{resource_exchange}.

The \texttt{PhysicsModule} base class handles work that is common to all \texttt{PhysicsModules}. It stores a reference to the main \texttt{Simulation} object and defines the \texttt{publish\_resource} function which is called during simulation setup (in \texttt{prepare\_simulation}) to handle the sharing of data. 

\subsubsection{The ComputeTool base class}
\texttt{ComputeTool} objects provide low level numerical functionality that is accessible to all physics modules. \texttt{PhysicsModules} and \texttt{ComputeTools} are somewhat similar. While \texttt{ComputeTool} is intended to be heavy on computations and light on data ownership, \texttt{PhysicsModule} is intended to own and manage data while delegating heavy computations to \texttt{ComputeTools}. An example of a compute tool could be a matrix equation solver that implements a specific algorithm. This solver could be called by the any physics module as needed on every time step. Another example could be a compute tool that constructs a linear interpolator for a given dataset. In this case, the work of the compute tool could happen during the initialization of the physics module that needs an interpolating function.

\subsubsection{The Diagnostic base class}
The diagnostics in turboPy are designed to separate the diagnostic and data output functionality from computations that happen in the \texttt{PhysicsModules}. At each timestep in the main simulation loop, every diagnostic gets a chance to examine the state of the simulation. The function \texttt{diagnose} must be defined for every subclass of the \texttt{Diagnostic} class. This function is the one that gets called at every timestep, and is where the class can do computations, output data, update plots, etc. The member function \texttt{finalize} can optionally be overridden so that the diagnostic can do any work that remains at the end of the simulation. This could include writing data buffers to file, closing any open file handles, etc.

\subsection{Detailed description of the resource exchange process}
\label{resource_exchange}
In computational physics codes written in frameworks like turboPy, the loose coupling between classes (provided by the use of physics modules) can have one main problem. Specifically, if one physics module needs to know about data from another physics module (e.g., if a module that handles charged particles needs the values of electromagnetic fields), then how can they communicate that data to each other in a systematic way? 

In turboWAVE this problem is solved using a method that has some similarities to the Publish-Subscribe, or Observer design pattern (see \cite[chp.~5]{gamma1994design} or \cite[chp.~2]{Freeman:2004:HFD:1076324}). TurboPy uses the same pattern that is used in turboWAVE, which facilitates the conversion of a turboPy module into a turboWAVE module. Examining this data sharing process will motivate the design of the resource sharing system used by the \texttt{Simulation} and \texttt{PhysicsModule} classes.

In turboPy, the class design used for data sharing is implemented within the \texttt{Simulation} and \texttt{PhysicsModule} classes. The problem that is being addressed with this design is that every physics module needs to both share resources and look for needed resources that are shared by other physics modules. The \texttt{Simulation} class keeps track of the list of instantiated physics modules, so that each module does not need to maintain a list of all other modules. The resource sharing process is set up at the start of the simulation by giving each PhysicsModule a chance to save a reference to the resources shared by other PhysicsModules. Then at each time step in the main loop, each physics module can use that reference to access the shared resources as needed. In pseudocode, the sharing setup process happens like this:
\begin{lstlisting}
add all physics modules to sim.physics_modules list
for physics module m in sim.physics_modules:
    for resource r that m wants to share:
        for physics module n in sim.physics_modules:
            if physics module n needs resource r:
                save a pointer to r in physics_module n
\end{lstlisting}
This complicated nested loop is cleanly broken apart in turboPy by defining several member functions for \texttt{PhysicsModules}. The goal of simplifying this loop is to separate as cleanly as possible the implementation (the nested loops) from the desired behavior (sharing data resources). This greatly simplifies the work required from the programmer to implement data sharing when writing a new \texttt{PhysicsModule}. The programmer specifies the desired behavior (what resource to provide, or share, with other physics modules, and what shared resources from other physics modules are needed) through two functions, \texttt{exchange\_resource} and \texttt{inspect\_resource}. The rest of the code that actually sets up the sharing is abstracted away, greatly simplifying the resource sharing code that needs to be written for each new PhysicsModule. 

The \texttt{Simulation} class takes care of the first step and the outer loop, so that we have something like this pseudocode in the \texttt{Simulation} class:
\begin{lstlisting}
add all physics modules to sim.physics_modules list
for physics module m in sim.physics_modules:
    m.exchange_resources()
\end{lstlisting}
where the new function \texttt{exchange\_resources} does this:
\begin{lstlisting}
m.exchange_resources() = 
    for resource r that m wants to share:
        for physics module n in sim.physics_modules:
            if physics module n needs resource r:
                save a pointer to r in physics module n
\end{lstlisting}
This is further simplified by extracting out the sharing that is being done in the inner loop. The \texttt{exchange\_resource} function becomes
\begin{lstlisting}
m.exchange_resources() = 
    for resource r that m wants to share:
        publish_resource(r)
\end{lstlisting}
Note that this pseudocode has the \texttt{publish\_resource(r)} statement inside a loop over resources \texttt{r}. It is more likely in any particular example of the \texttt{exchange\_resource} function that there will not be a loop, but simply a sequence of \texttt{publish\_resource} statements, one for each resource that is to be shared. 

The \texttt{publish\_resource} function should then have a form such as this:
\begin{lstlisting}
publish_resource(r) = 
    for physics module n in sim.physics_modules:
        if physics module n needs resource r:
            save a reference to r in physics module n
\end{lstlisting}
One final simplification is to abstract out the work of checking the resource and saving it if it is needed:
\begin{lstlisting}
publish_resource(r) = 
    for physics module n in sim.physics_modules:
        n.inspect_resource(r)
\end{lstlisting}
where
\begin{lstlisting}
n.inspect_resource(r) = 
    if physics module n needs resource r:
        save a reference to r in physics module n
\end{lstlisting}
Note that in order for this simplification process to work, the physics modules need to know about their parent or owner \texttt{Simulation} class instance.

By abstracting the sharing loop in this way, newly written \texttt{PhysicsModules} only need to override the definitions of the two functions \texttt{exchange\_resources} and \texttt{inspect\_resource}. The function \texttt{exchange\_resources} is where the new \texttt{PhysicsModule} describes data or resources that it wants to share, and the function \texttt{inspect\_resource} is where the new \texttt{PhysicsModule} looks for shared data that it needs.

In the turboPy \texttt{PhysicsModule} base class, there are stubs for each of these functions. Any subclass that wants to use this data sharing functionality needs only to override these two functions. If the physics module has data to share, the function \texttt{exchange\_resources} will look something like this:
\begin{lstlisting}[language=python]
def exchange_resources(self):
    self.publish_resource({"MyPhysicsModule:data": self.data})
\end{lstlisting}
This uses the function \texttt{publish\_resource} described in pseudocode above. In turboPy the actual code is nearly as simple:
\begin{lstlisting}[language=python]
def publish_resource(self, resource: dict):
    for physics module in self.owner.physics_modules:
        physics_module.inspect_resource(resource)
    for diagnostic in self.owner.diagnostics:
        diagnostic.inspect_resource(resource)
\end{lstlisting}
Here the string \texttt{description} is used to identify the resource being shared. While there are no rules about the form this string should take, it often takes a form that combines the name of the physics module that provides the resource with the resource name. For example, a \texttt{description} string for the charge/current density four-vector field provided by an \texttt{Electromagnetic} physics module might be \texttt{electromagnetic:sources}.

If a custom \texttt{PhysicsModule} needs resources shared by other physics modules, it can override the \texttt{inspect\_resource} function. For example:
\begin{lstlisting}[language=python]
def inspect_resource(self, resource):
    if "MyPhysicsModule:data" in resource:
	self.data = resource["MyPhysicsModule:data"]
\end{lstlisting}
In this example, the physics module is looking for data that has been shared with the identifier \texttt{MyPhysicsModule:data}. If it is found, then the physics module saves it in the member variable \texttt{self.data}.

\section{Implementation details for turboPy}
\label{turbopy}

The implementation of the class design described in Sec.~\ref{design} is fairly straightforward in turboPy. Most of the implementation can be understood fairly easily just by reading the source code, but two features in particular deserve some additional explanation.
First, the sharing of data between physics modules is achieved using mutable python variables, and this is described in Sec.~\ref{mutable}.
The second feature is the use of subclass ``registries'' to create libraries of physics modules, compute tools, and diagnostics in turboPy. This provides a simple mechanism for extending the capabilities of computational physics codes written within the turboPy framework---adding custom subclasses to the appropriate class registry. This process is described in Sec.~\ref{registry}.

\subsection{Sharing resources with mutable python variables}\label{mutable}
In turboWAVE, the resource sharing implemented in the \texttt{Module} method \texttt{Inspect\-Resource} is based on being able to use C++ pointers to share a memory address. When different turboWAVE modules need access to the same underlying data set, they can use these shared pointers to access the data. Python does not have pointers, so the question in python becomes how to store a reference to the shared data. In python, everything is an object, and using the assignment operator ``='' can be thought of as giving a name to an already existing object.
Because everything is an object, there is no problem giving multiple names to the same object. So sharing a resource between physics {modules} is as simple as saving another ``name'' for the object:
\begin{lstlisting}[language=Python]
physics_module1.data = 1
physics_module2.data = physics_module1.data
# After this, physics_module1.data == physics_module2.data
\end{lstlisting}
This is similar in many ways to thinking of the variables as if they ``are'' pointers.
There is a subtlety that arises, however, especially for programmers not used to python. In python, some objects are mutable, meaning their value can change, and some are immutable, meaning their value cannot change. In the above example, the object that is named \texttt{physics\_module1.data} is the integer 1, and integers in python are immutable. If you try to ``change'' its value by using assignment, you may be surprised by the behavior.
\begin{lstlisting}[language=Python]
# Try to change the value of physics_module1.data
physics_module1.data = 2
\end{lstlisting}
After this statement, \texttt{physics\_module1.data == physics\_module2.data} is no longer true. The ``pointer'' \texttt{physics\_module1.data} is now pointing to a new object, the python integer 2. 

Because python works this way, care must be taken when trying to share data between physics modules. First, the resources that physics modules share should be mutable python objects, such as lists or numpy arrays. This way, changes that are made in one physics module can be seen by the other modules that are sharing the resource. Second, care should be taken that no statements cause the variable name to ``point'' to a different object. This is perhaps best illustrated by an example. First create two variables that point to the same mutable object (a python list):
\begin{lstlisting}[language=Python]
a = [1]
b = a
\end{lstlisting}
After these lines, both variables point to the same python list, which has only one element, the integer 1. If we now want to now change the value of data using the variable \texttt{a}, we must use care. It might be tempting to do this:
\begin{lstlisting}[language=Python]
a = [2]
\end{lstlisting}
However, \texttt{a} is now pointing to a {\em different} python list. It is no longer true that \texttt{a~==~b}. Instead, we should do an operation that mutates the values in the existing list, by using indexing:
\begin{lstlisting}[language=Python]
a[0] = 2
\end{lstlisting}
This statement changes the value of the existing list, and both \texttt{a} and \texttt{b} still point to the same object. The change made by this line (which mutates the object that \texttt{a} points to) will be reflected in \texttt{b}.\footnote{A useful tool for tracking down problems related to mutating an object versus changing which object a variable is pointing to is the built-in python function \texttt{id}. This returns a unique identifier, which in CPython is actually the object's memory address.}

\subsection{PhysicsModule, ComputeTool, and Diagnostic: Factory classes with dynamic registration}\label{registry}
One common problem addressed by the turboPy framework is the translation of an input definition (from a file, a configuration dictionary, or other problem definition) into concrete class objects for custom \texttt{PhysicsModules}, \texttt{Diagnostics}, and \texttt{ComputeTools}. In turboPy, this problem is solved by using the factory pattern \cite[chp.~3]{gamma1994design}, but with an extendable registry of concrete classes. This extendable factory functionality is provided in turboPy via a simple base class called \texttt{DynamicFactory}, from which the \texttt{PhysicsModule}, \texttt{Diagnostic}, and \texttt{ComputeTool} classes are derived. This base class provides a dictionary of registered subclasses, and two class methods (\texttt{register} and \texttt{lookup}) that interrogate the dictionary.
\begin{lstlisting}[language=Python]
class DynamicFactory:
    """
    This base class provides a dynamic factory pattern functionality to classes that derive from this.
    """
    _factory_type_name = "Class"
    
    @classmethod
    def register(cls, name_to_register, class_to_register):
        if name_to_register in cls._registry:
            raise ValueError("{0} '{1}' already registered".format(cls._factory_type_name, name_to_register))
        cls._registry[name_to_register] = class_to_register

    @classmethod
    def lookup(cls, name):
        try:
            return cls._registry[name]
        except KeyError:
            raise KeyError("{0} '{1}' not found in registry".format(cls._factory_type_name, name))

    @classmethod
    def is_valid_name(cls, name):
        return name in cls._registry
\end{lstlisting}
As can be seen in the above implementation of the \texttt{DynamicFactory} class, python class methods are provided that allow the class definitions of any \texttt{DynamicFactory} subclass to be modified at runtime. Thus the base python class itself becomes the dynamic factory, without requiring an instance of the class.

Because of this extendable factory design, new functionality for turboPy can be easily added by users. For example, a new \texttt{PhysicsModule} subclass can be written and added to turboPy by something like this code:
\begin{lstlisting}[language=Python]
from turbopy import Simulation, PhysicsModule
class MyPhysicsModule(PhysicsModule):
	def __init__(...):
		# Custom initialization code here
		
	def update(self):
		# Custom update code here

PhysicsModule.register("MyPhysicsModule", MyPhysicsModule)
\end{lstlisting}
Note that \texttt{MyPhysicsModule} is now useable by the \texttt{Simulation} class, but no code within the turboPy package itself needs to be modified in order to do so. This is a significant usability advantage in the turboPy package, because it allows developers to quickly add new functionality to turboPy via new \texttt{PhysicsModules}, \texttt{ComputeTools}, and \texttt{Diagnostics}. The details about how to make all of the new code work with the rest of the framework is abstracted away, and the developer is freed to focus on their specific physics, algorithm, or diagnostic. Using the new \texttt{PhysicsModule} is achieved through looking up the class in the \texttt{PhysicsModule} registry:
\begin{lstlisting}[language=Python]
physics_module_class = PhysicsModule.lookup("MyPhysicsModule")
physics_module_instance = physics_module_class(owner=sim, input_data=instance_data)
\end{lstlisting}
In this code, a new instance of a physics module of type \texttt{MyPhysicsModule} is created with the instance \texttt{sim} of the \texttt{Simulation} class as the owner and the specific physics module parameters provided in the dictionary \texttt{instance\_data}.

\section{Example of turboPy usage}
\label{example}

Because of the flexible design and modular structure of turboPy, there are a wide range of possible use cases for this framework. Using turboPy, it is straightforward to translate a high-level workflow into specific turboPy physics modules and diagnostic subclasses. The example in this section illustrates this process for the specific example of the motion of a charged particle in an electric field {\cite{example-app}}. This simple example lays some of the groundwork that would be needed for a full particle-in-cell (PIC) simulation code. Some features that this example has in common with a PIC code are: 
\begin{itemize}
\item field stored on a grid, 
\item interpolation of the field to the particle location, 
\item use of a {\tt ComputeTool} for defining a ``particle pusher'', 
\item and  output of particle and field diagnostics at each time-step of a simulation clock.
\end{itemize}

The motion of a charged particle in an electromagnetic field is governed by Maxwell's equations and the Lorentz force equation. Here we only consider electric fields that are right-moving waves, which satisfy
\begin{eqnarray}\label{field}
\partial_t E_y = -c \partial_x E_y.
\end{eqnarray}
In this case, the electric field is only in the $y$-direction, and only varies with respect to $x$ and $t$.

The particle is taken to be non-relativistic, and thus its equation of motion is (Jackson Eq.~11.124 \cite{jackson}, with ${\bf p} = m{\bf v}$):
\begin{eqnarray}
\frac{d{\bf v}}{dt} = \frac{q}{m}\left( {\bf E} + {\bf v} \times {\bf B} \right).
\end{eqnarray}
To simplify this equation, we will neglect the magnetic field.  

\begin{figure}[ht]
\begin{center}
\includegraphics[width=7cm]{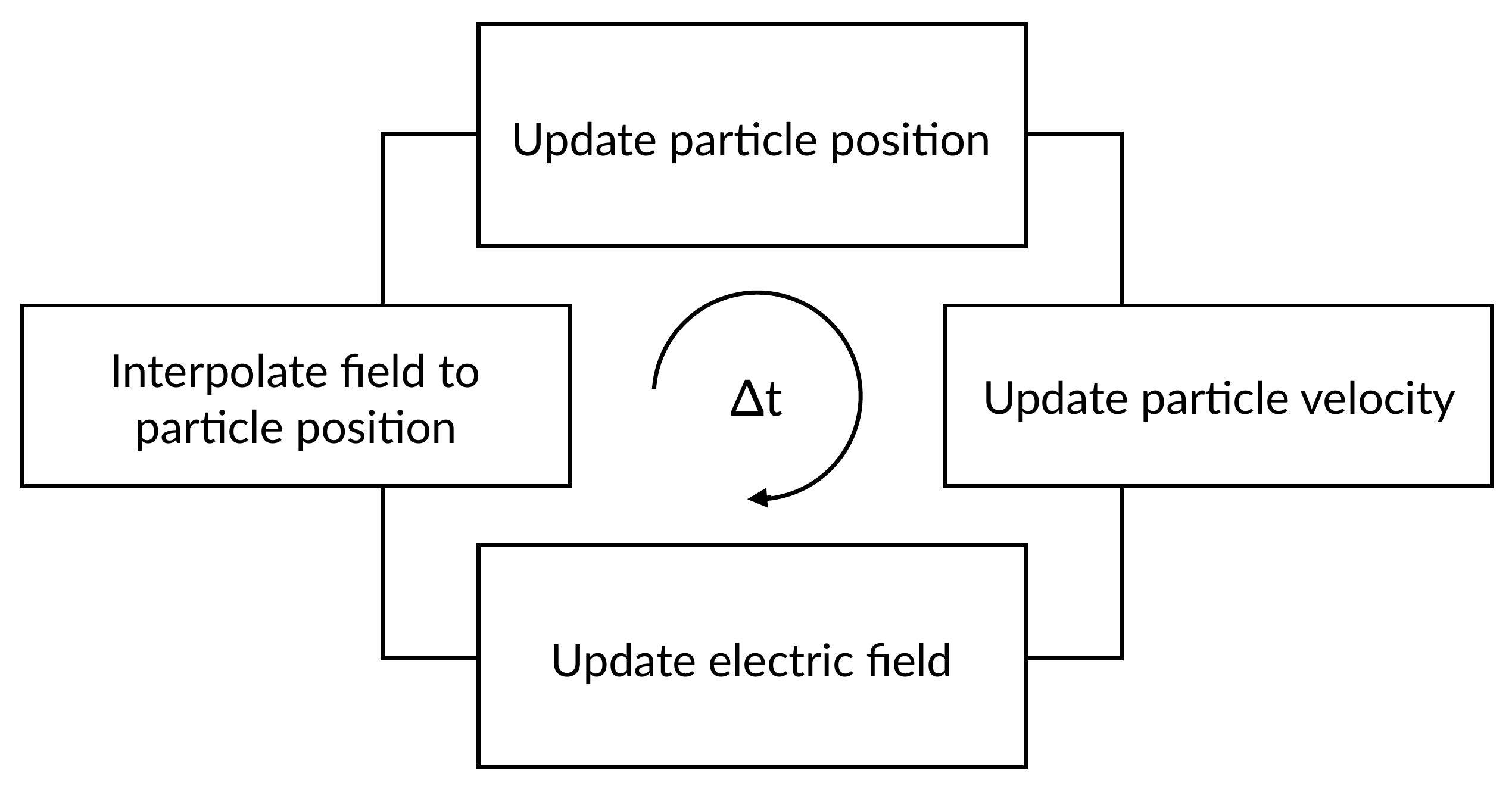}
\caption{Workflow for solving equations of motion of a charged particle in an electric field.}
\label{ex:particle}
\end{center}
\end{figure}

A particle initially at rest will only move in the $y$-direction, with motion described by:
\begin{eqnarray}
\frac{dy}{dt} &=& v_y, \label{part1}\\
\frac{dv_y}{dt}  &=& \frac{q}{m} E_y. \label{part2}
\end{eqnarray}
In order to solve these equations, an iterative workflow will be adopted, as illustrated in Fig.~\ref{ex:particle}. This workflow shows the computations that need to be performed at each time step: update the particle position and velocity (using Eqs.~\ref{part1} and \ref{part2}), update the electric field (using Eq.~\ref{field}), and then interpolate the new field values to the location of the particle.

\begin{figure}[htb]
\begin{center}
\includegraphics[width=9cm]{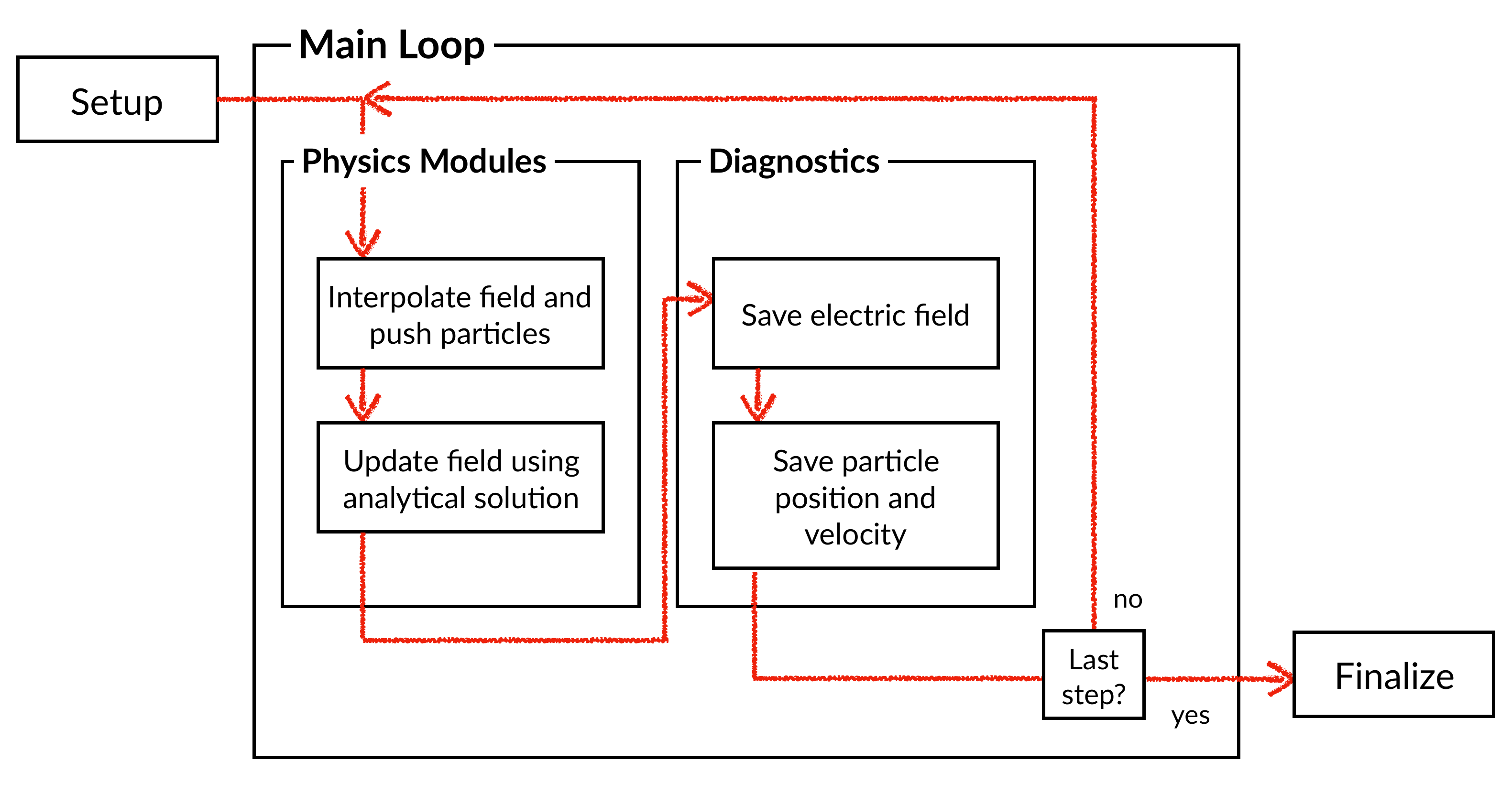}
	\caption{Diagram illustrating one possible implementation of the example workflow using custom turboPy physics modules and diagnostics. Each box in the flow represents a PhysicsModule or Diagnostic subclass written to perform the described action. }
\label{ex:particle_modules}
\end{center}
\end{figure}
One possible implementation of this workflow as turboPy physics modules is shown in Fig.~\ref{ex:particle_modules}. In this implementation, the workflow is divided between two custom turboPy physics modules. The first physics module (named {\tt ChargedParticle} in the source) takes care of updating the particle position and velocity. Since interpolating the field value to the particle position is closely related to the particle update, it is also taken care of in this physics module. The update of the particle's position and momentum is performed through the use of a custom {\tt ComputeTool}. This tool, called {\tt ForwardEuler}, defines a function {\tt push} which updates the particle based on the value of the electric field. By separating out the implementation details of the particle push into a separate {\tt ComputeTool}, the code becomes more flexible, allowing for the substitution of a different update algorithm at a later time. For example, if the magnetic field were included then it would make sense to use the Boris algorithm to push the particle.

The second physics module (named {\tt EMWave} in the source) updates the value of the electric field. In this example, an analytical solution of Eq.~\ref{field} has been implemented:
\begin{eqnarray}
E_y(x,t) = E_0 \cos(2\pi (kx - \omega t)).
\end{eqnarray}
Here $\omega$ is a free parameter, and $k$ is computed from $k = \omega/c$. At each time step, the {\tt EMWave} physics module evaluates this solution for each point on the grid. Note that separating  the electric field update from the particle dynamics allows for simple modifications to this code for solving different types of problems. It is easy to imagine other physics modules which implement different types of fields; these could be other types of waves, fields driven by circuits, etc.

In addition to the custom physics modules, a custom diagnostic class, {\tt ParticleDiagnostic}, is defined. This diagnostic class saves a reference to the particle's position (or momentum). At each time step, it accumulates the position (or momentum) into a buffer, which is written to file at the end of the simulation. 
Since the {\tt EMWave} physics module uses the built-in turboPy {\tt field} functionality to store the electric field on the grid, it also uses the built-in {\tt FieldDiagnositc} class to save the electric field values to file. No new diagnostic class was needed.

\begin{figure}[htbp]
\begin{center}
\includegraphics[width=6cm]{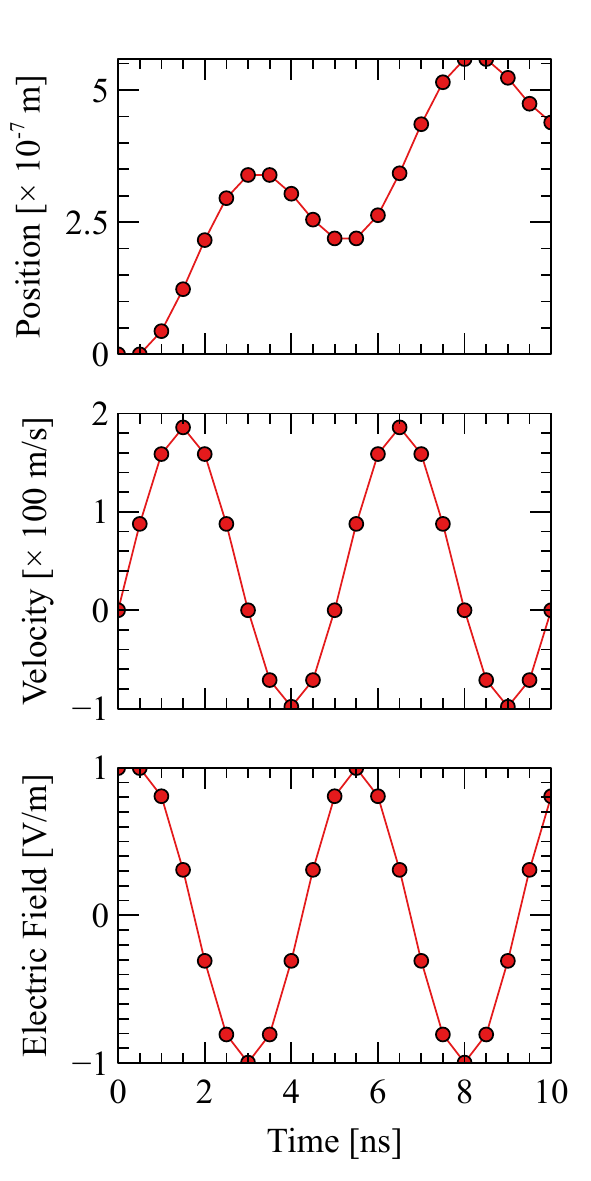}
\caption{Output from the example problem, showing the motion of a test particle in an electromagnetic wave.}
\label{example_plots}
\end{center}
\end{figure}
This example problem is implemented in two files: a python source code file {\tt particle\_in\_field.py} (listed in \ref{example_source}) and a plain-text configuration file {\tt particle\_in\_field.toml} (listed in \ref{example_config}). {These files are also available from the GitHub repository \cite{example-app}.} The python file is where the custom \texttt{PhysicsModule} and \texttt{Diagnostic} classes are defined, and then added to the \texttt{PhysicsModule} and \texttt{Diagnostic} registries. The parameters for the simulation we want to run are then read in from the configuration file\footnote{The TOML format\cite{toml} is used for the configuration file for several reasons. It is a simple plain-text format and so is easy to create by hand. It is flexible, but with a well defined format specification. There are several python libraries which make reading TOML files simple; turboPy uses {\tt qtoml}\cite{qtoml}.}, and converted into a python dictionary. Alternatively, this dictionary could have been created directly in python without the need for an input file. This configuration dictionary is used to create and run a \texttt{Simulation} object.
Figure~\ref{example_plots} shows the outputs of the simulation, which were saved to file by the field and particle diagnostics.

\section{Conclusions}
\label{con}
A new, lightweight framework called turboPy{\cite{turbopy}} has been written in python for streamlining any computational physics workflows that requires a grid and a clock. This includes everything from a full plasma simulation code to a quick custom post processor for the results of an entirely separate code. It is designed to mirror the class structure used in the code turboWAVE, to simplify the translation of turboPy modules to turboWAVE. This class design provides an easily extendable framework in which new physics modules can be quickly developed and tested.

The example shown in Sec.~\ref{example} illustrates how to use turboPy for a real (if simplistic) physics problem. The equations defining the problem were translated into a flowchart showing the work that needs to be done at each time step. This workflow was then broken up into two sections which then translated directly into two custom turboPy \texttt{PhysicsModule} classes. This same process---physics model to workflow to turboPy \texttt{PhysicsModule} {subclasses}---can be used for many different types of problems. 

By using the framework provided by turboPy, a computational scientist can reduce their cognitive burden while designing a new code to implement a desired workflow. If the workflow can be sketched out as a process that happens repetitively in a time loop, then translating that sketch into a turboPy block diagram is quite straightforward. From there, writing the custom \texttt{PhysicsModule} and \texttt{Diagnostic} subclasses which implement the block diagram is a fairly well-defined task: each block translates nearly one-to-one into a subclass which does the work for that block.

There are many ways in which turboPy can be enhanced and extended. These range from creating a GUI for developing turboPy block diagrams and problem setup to providing a python package so that turboPy can be installed via package installers. Also, the existing grid and clock options are currently fairly rudimentary, and it is straightforward to imagine how multidimensional grids, finite differencing tools, and more could be added. {Linking between turboPy and turboWAVE could potentially be achieved using a C++/Python bridging tool such as \texttt{pybind11}. However, t}hese efforts are left for future work, or possibly separate add-on projects in order to keep the core turboPy framework as lightweight as possible.

\section{Acknowledgements}

Funding: This work was supported by the Naval Research Laboratory base program.



\appendix
\section{TurboPy Example Source}\label{example_source}

{This example turboPy app is available for download from GitHub. \cite{example-app}}

\begin{lstlisting}[language=Python]
from turbopy import Simulation, PhysicsModule, Diagnostic, CSVDiagnosticOutput, ComputeTool
import numpy as np


class EMWave(PhysicsModule):
    def __init__(self, owner: Simulation, input_data: dict):
        super().__init__(owner, input_data)
        self.c = 2.998e8
        self.E0 = input_data["amplitude"]
        self.E = owner.grid.generate_field()
        self.omega = input_data["omega"]
        self.k = self.omega / self.c
    
    def initialize(self):
        phase = - self.omega * 0 + self.k * (self.owner.grid.r - 0.5)
        self.E[:] = self.E0 * np.cos(2 * np.pi * phase)

    def update(self):
        phase = - self.omega * self.owner.clock.time + self.k * (self.owner.grid.r - 0.5)
        self.E[:] = self.E0 * np.cos(2 * np.pi * phase)
        
    def exchange_resources(self):
        self.publish_resource({"EMField:E": self.E})


class ChargedParticle(PhysicsModule):
    def __init__(self, owner: Simulation, input_data: dict):
        super().__init__(owner, input_data)
        self.E = None
        self.x = input_data["position"]
        self.interp_field = owner.grid.create_interpolator(self.x)
        self.position = np.zeros((1, 3))
        self.momentum = np.zeros((1, 3))
        self.eoverm = 1.7588e11
        self.charge = 1.6022e-19
        self.mass = 9.1094e-31
        self.push = owner.find_tool_by_name("ForwardEuler").push
        
    def exchange_resources(self):
        self.publish_resource({"ChargedParticle:position": self.position})
        self.publish_resource({"ChargedParticle:momentum": self.momentum})
    
    def inspect_resource(self, resource):
        if "EMField:E" in resource:
            self.E = resource["EMField:E"]    

    def update(self):
        E = np.array([0, self.interp_field(self.E), 0])
        self.push(self.position, self.momentum, self.charge, self.mass, E, B=0)


class ParticleDiagnostic(Diagnostic):
    def __init__(self, owner: Simulation, input_data: dict):
        super().__init__(owner, input_data)
        self.data = None
        self.component = input_data["component"]
        
    def inspect_resource(self, resource):
        if "ChargedParticle:" + self.component in resource:
            self.data = resource["ChargedParticle:" + self.component]
    
    def diagnose(self):
        self.output_function(self.data[0, :])

    def initialize(self):
        # setup output method
        functions = {"stdout": self.print_diagnose,
                     "csv": self.csv_diagnose,
                     }
        self.output_function = functions[self.input_data["output_type"]]
        if self.input_data["output_type"] == "csv":
            diagnostic_size = (self.owner.clock.num_steps + 1, 3)
            self.csv = CSVOutputUtility(self.input_data["filename"], diagnostic_size)

    def finalize(self):
        self.diagnose()
        if self.input_data["output_type"] == "csv":
            self.csv.finalize()

    def print_diagnose(self, data):
        print(data)

    def csv_diagnose(self, data):
        self.csv.append(data)


class ForwardEuler(ComputeTool):
    def __init__(self, owner: Simulation, input_data: dict):
        super().__init__(owner, input_data)
        self.dt = None
        
    def initialize(self):
        self.dt = self.owner.clock.dt
    
    def push(self, position, momentum, charge, mass, E, B):
        p0 = momentum.copy()
        momentum[:] = momentum + self.dt * E * charge
        position[:] = position + self.dt * p0 / mass


PhysicsModule.register("EMWave", EMWave)
PhysicsModule.register("ChargedParticle", ChargedParticle)
Diagnostic.register("ParticleDiagnostic", ParticleDiagnostic)
ComputeTool.register("ForwardEuler", ForwardEuler)

input_file = "particle_in_field.toml"
sim = construct_simulation_from_toml(input_file)
sim.run()
\end{lstlisting}

\section{TurboPy Example Problem Configuration}\label{example_config}
\begin{lstlisting}
# Define simulation related parameters
[Grid]
N = 30
r_min = 0.0
r_max = 1.0

[Clock]
start_time = 0.0
end_time = 1e-8
num_steps = 20


# Add the physics modules
[PhysicsModules]         
[PhysicsModules.EMWave]
amplitude = 1.0
omega = 2e8

[PhysicsModules.ChargedParticle]
position = 0.5


# Add the compute tools
# Empty "heading" is ok, adds the tool with no/default parameters
[Tools.ForwardEuler]


# Add the diagnostics
[Diagnostics]
# First add some default diagnostic parameters
directory = "output/"
output_type = "csv"

[Diagnostics.grid]
filename = "grid.csv"

[Diagnostics.clock]
filename = "time.csv"

[[Diagnostics.field]] # Double brackets here because we can have multiples of these...
component = 0
field = "EMField:E"
filename = "efield.csv"

[Diagnostics.point] # single brackets here because we only have one of these...
field = "EMField:E"
location = 0.5
filename = "e_0.5.csv"

[[Diagnostics.ParticleDiagnostic]]
component = "momentum"
filename = "particle_p.csv"

[[Diagnostics.ParticleDiagnostic]]
component = "position"
filename = "particle_x.csv"
\end{lstlisting}









\end{document}